\documentclass{aa}  

\usepackage{graphicx}
\usepackage{txfonts}
\begin{document}

\title{Photometric determination of the mass accretion rates of pre-main-sequence stars}

\subtitle{ IX. Recent star formation in the periphery of NGC 346}

\author{C. Dik
      \inst{1}
      \and
      G. De Marchi\inst{2}\fnmsep
      }

\institute{Leiden Observatory, Leiden University, 2300 RA Leiden, The Netherlands\\
          \email{dik@strw.leidenuniv.nl}
     \and
         European Space Research and Technology Centre, Keplerlaan 1, 2200 AG Noordwijk, The Netherlands\\
         \email{gdemarchi@esa.int}
}

\titlerunning{Recent star formation in the periphery of NGC 346}         

\date{Received ; accepted }

 
  \abstract 
   {We  {studied} the properties of star formation and the characteristics of young stars in a quiet region located beyond the outskirts of the prominent star-forming cluster NGC 346 in the Small Magellanic Cloud (SMC). Utilising observations from the \textit{Hubble} Space Telescope across the broad $V$ and $I$ bands, as well as the narrow $H\alpha$ band, we identified populations with ages of roughly 10, 60, 400~Myr, and 5~Gyr through isochrone comparison. We successfully identified 137 bona fide pre-main-sequence (PMS) candidates exhibiting H$\alpha$ excess with a significance level of 5$\sigma$ accompanied by an H$\alpha$ line emission equivalent width exceeding 20\,$\mathrm{\AA}$. Physical parameters for these PMS stars were determined, including mass, age, accretion luminosity, and mass accretion rate. Most PMS stars have an age of around 16 Myr and an average mass of $0.80 \pm 0.16$\,M$_\odot$. The median mass accretion rate for all 137 PMS stars is estimated to be $\Dot{M}_{\rm acc} \simeq 8.0 \times 10^{-9} \mathrm{~M{_\odot}yr^{-1}}$. While this rate is lower than that observed in the NGC 346 cluster itself, it is comparable with those measured for PMS stars in low-density star-forming regions in the SMC, despite the absence of apparent clustering and nebulosity. Furthermore, our analysis reveals that the ratios of accreting and non-accreting PMS stars to non-PMS stars and their mass accretion rate correlate with their distance from a group of hot massive stars in the vicinity. This suggests that the ultraviolet radiation emitted by these massive stars might erode the circumstellar discs of nearby PMS stars. Lastly, the overlap between our studied region and observations from the \textit{James Webb} Space Telescope reveals that some of the identified PMS stars display near-IR excess.
   }


\maketitle

\section{Introduction}

The process of star formation is a complex and fundamental phenomenon in astrophysics, and is intricately connected to the evolution of galaxies and star systems. It begins with the collapse of gas clouds; this leads to the formation of protostars, which eventually evolve into stable main sequence (MS) stars --- the predominant stage of a star's life, including our own Sun \citep{starformationreview2007ARA&A..45..565M}. Models of star formation have been refined over time, incorporating various physical factors such as turbulence \citep{sf1987ARA&A..25...23S}, magnetic fields \citep{magneticsf1999ApJ...520..706C}, and radiation \citep{windsradiationsf2012ApJ...754...71K}. 

The environment plays a critical role in shaping the process of star formation. Factors such as gas density, temperature, and metallicity exert significant  {influence} on the fragmentation and subsequent properties of forming stars \citep{envsf2023MNRAS.519.5017N}. High-density environments tend to favour the formation of more massive stars, while the presence of young massive stars can either stimulate or hinder the star formation process through their winds and radiation \citep{windradsf2018MNRAS.478.4799H}. Similarly, variations in metallicity are known to affect the rates of mass accretion, thereby influencing the evolution of star-forming regions \citep{metallicity30dor2017ApJ...846..110D,demarchietal2024}. The conditions of low-metallicity regions are more comparable to those of the early Universe and thereby provide useful information on the star formation process in the past \citep[see e.g.][]{maddenetal2013, bromm2013}.  {Furthermore, studying star formation outside the boundaries of a massive cluster offers a unique opportunity to test how far the influence of clustered, high‑mass star formation extends into the surrounding lower-density field.}

{In this work we probe the conditions mentioned above by studying a field in the periphery of the NGC\,346 starburst cluster in the Small Magellanic Cloud (SMC). This field has a low metallicity, a low density, and is close to a massive cluster, conditions thought to be common in the early Universe. Such environments are known to respond differently to radiative and mechanical feedback {compared to} more metal‑rich, higher‑density star‑forming regions \citep[see e.g.][]{krumholz2012,remyruyeretal2014}. We searched for, identified, and characterised accreting pre‑main-sequence (PMS) stars, investigated their physical properties, and assessed how the local environment shapes the star formation process in this diffuse region.}

The structure of the paper is as follows: In Section \ref{sec:observations} we present the data and observations. Section \ref{sec:photometry} describes the photometric analysis of the data, including the colour-magnitude diagram (CMD) and the necessary corrections. In Section \ref{sec:pms} we discuss how to identify PMS stars and present the properties of the PMS population in the studied region. In Section \ref{sec:nearby_stars} we derive the effect of nearby stars on star formation. Section \ref{sec:comparison} contains a comparison with recent \textit{James Webb} Space Telescope (JWST) observations in the region. Finally, a summary and conclusions are provided in Section \ref{sec:conclusion}.

\section{Observations}\label{sec:observations}

\begin{figure}
   \centering
   \includegraphics[width=8cm]{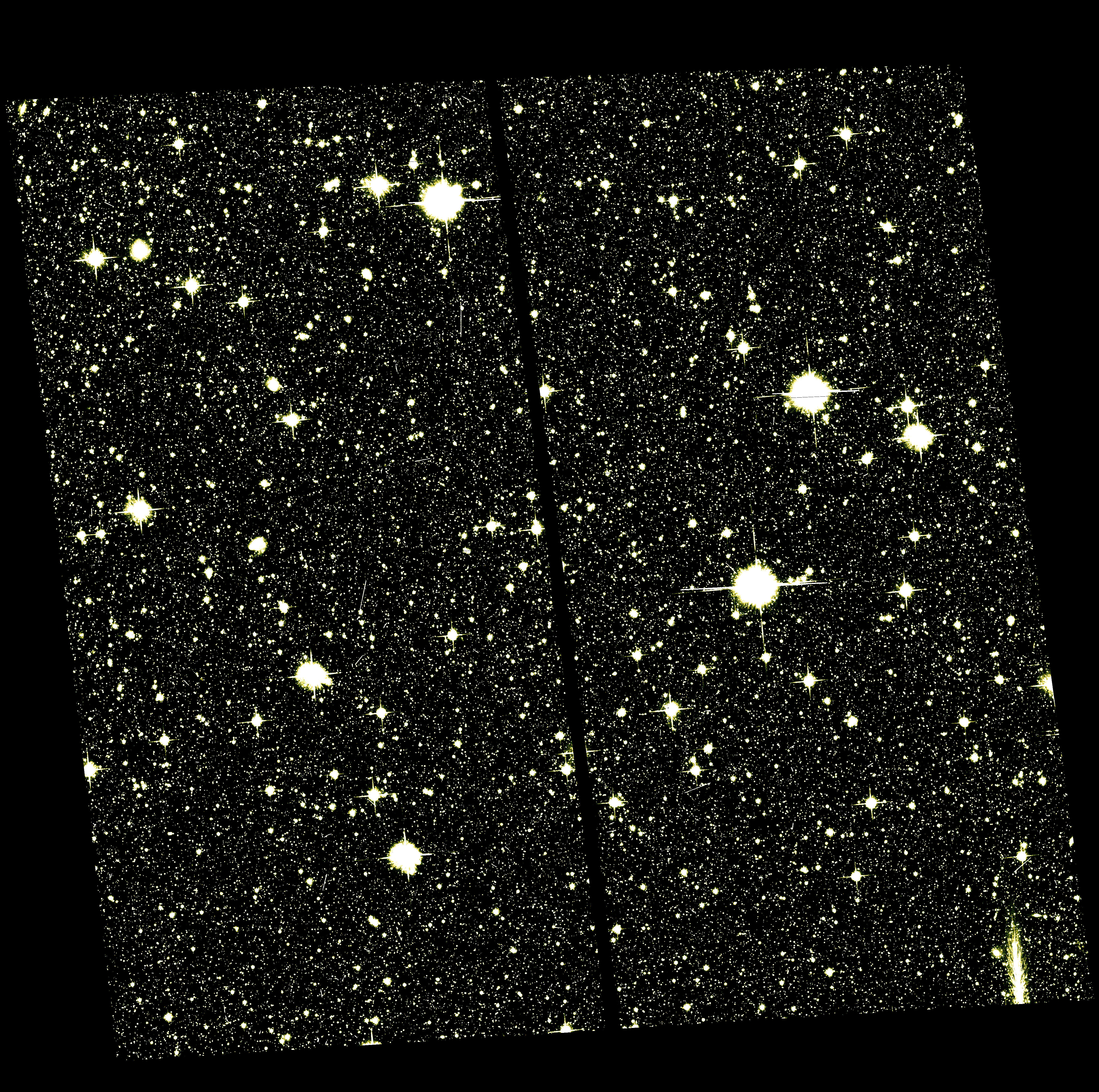}
      \caption{Colour image of the studied field south of NGC346, created by combining the HST $V$- and $I$-band images.
              }
         \label{fig:field}
\end{figure}

This work mainly uses observations collected with the \textit{Hubble} Space Telescope (HST), in the F555W band (hereafter `$V$ band'),  F814W band (`$I$ band'), and F656N band (`$H\alpha$ band'), with exposure times of 1200\,s, 400\,s and 2394\,s, respectively. The observations in the $V$ and $I$ band were executed with the Wide Field Channel on the Advanced Camera for Surveys (WFC/ACS) in July 2004 under the HST proposal ID 10248 (A. Nota). The narrow-band H$\alpha$ images were obtained with the Ultraviolet/Visible channel of the Wide Field Camera 3 (UVIS/WFC3), in August 2013 under the HST proposal ID 13009 (G. De Marchi). Both proposals have the goal of studying star formation in the SMC and characterising PMS stars \citep{nota2006ApJ...640L..29N}. The H$\alpha$ images covered regions with preexisting $V$ and $I$ observations, in order to identify PMS stars, and derive accretion luminosities and mass accretion rates of the studied objects directly from the photometry \citep{PMSngc346guido2011ApJ...740...11D}. The studied region can be seen in colour in Fig. \ref{fig:field}. The field spans $200\arcsec \times 200\arcsec$, corresponding to $\sim 60 \times  60$ pc at a distance of $62.5$ kpc \citep{distanceSMC2020ApJ...904...13G}, and lies to the south of the main cluster NGC 346. 

\begin{figure}
   \centering
   \includegraphics[width=9cm]{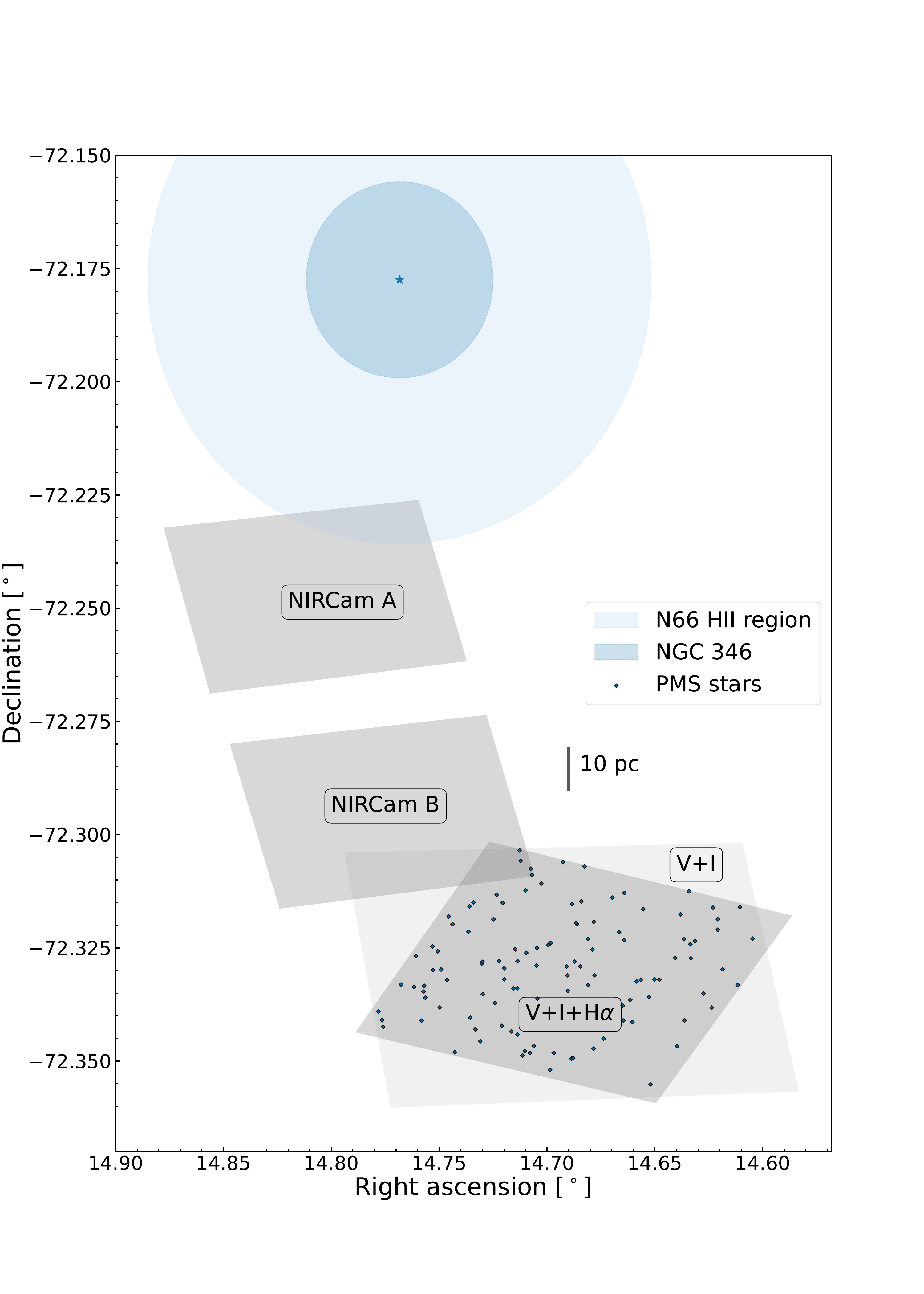}
      \caption{Regions observed with HST and JWST, and the rough extent of the cluster NGC 346 and the surrounding HII region N66.
              }
         \label{fig:fields}
\end{figure}

In addition to the HST images, this work uses JWST photometry of a nearby field to make comparisons with the studied region. This consists of observations using the NIRCam instrument in the F090W, F115W, F182M, F187N, F277W, F356W, F405N, and F430M filters. These observations were carried out as part of  {programme} 1227 (M. Meixner) in July 2022. The objective of the programme is a comprehensive imaging and spectroscopic study of the cluster NGC 346 \citep{Jones2023NatAs...7..694J, demarchietal2024}. During the spectroscopic observations of the cluster with NIRSpec, NIRCam was used to observe in parallel a field located about $10^\prime$ south of the cluster. The area covered in this way spans from the cluster edge southwards towards the field studied with HST. There is a small corner of the NIRCam field of view (B side of the camera) that overlaps with the $V$, $I,$ and $H\alpha$ images (top left in Fig. \ref{fig:field}). The relative positions of the HST and JWST fields --- including the overlapping region --- as well as the approximate extent of the NGC 346 cluster can be seen in Fig. \ref{fig:fields}. Table \ref{tab:imaging_data} lists the observations used in this  {work.}

\begin{table}[h]
\caption{List of observations.}
\begin{tabular}{llccc}
\hline \hline
{Camera}   & {Filter} & {Time [s]} & {RA} & {Dec} \\ \hline
 {WFC/ACS}   & F555W & 1200 & $14.6880$ & $-72.3307$ \\
 {WFC/ACS}   & F814W &  ~~400 & $14.6880$ & $-72.3307$ \\
 {UVIS/WFC3} & F656N & 2394 & $14.6880$ & $-72.3307$ \\
\hline
 {NIRCam} & F090W & ~~299  & $14.7917$ & $-72.2711$ \\
 {NIRCam} & F115W & ~~299  & $14.7917$ & $-72.2711$ \\
 {NIRCam} & F182M & ~~299  & $14.7917$ & $-72.2711$ \\ 
 {NIRCam} & F187N & 1353 & $14.7917$ & $-72.2711$ \\
 {NIRCam} & F277W & ~~299  & $14.7917$ & $-72.2711$ \\
 {NIRCam} & F356W & ~~299  & $14.7917$ & $-72.2711$ \\
 {NIRCam} & F405N & 1353 & $14.7917$ & $-72.2711$ \\
 {NIRCam} & F430M & ~~299  & $14.7917$ & $-72.2711$ \\
\hline
\end{tabular}
\tablefoot{
The observations obtained with the HST are listed in the top panel, those with JWST at the bottom. Exposure times are in second, right ascension (RA) and declination (Dec) in decimal degree.}
\label{tab:imaging_data}
\end{table}

\section{Photometric analysis}\label{sec:photometry}

The $V$, $I$, and $H\alpha$ images were retrieved from the HST archive and had already been processed through the standard calibration pipelines for the ACS ($V$ and $I$ bands) and WFC3 ($H\alpha$ band) instruments, which account for bias subtraction, flat‑fielding, dark correction, and image drizzling.  {We used the \texttt{DAOPHOT} routine in \texttt{IRAF} on all images to perform the source detection and photometry, running the detection step on the $I$‑band image because it is the deepest.}  {Sources were selected using an aperture radius of 2 pixels,} and  {we required their peak flux to exceed the local background by at least $5\,\sigma$.} The background was measured  {in an annulus between 3 and 5 pixels from the source centre, with $\sigma$ defined as the standard deviation of the counts within this annulus.}  {Photometric uncertainties were computed with the \texttt{DAOPHOT} routine from photon‑noise statistics of the source and from the background variance within the aperture, and were propagated into magnitude uncertainties using standard error‑propagation formulas.}  {This procedure yielded 42\,299 sources in the $I$‑band image, of which 42\,168 have counterparts detected in the $V$ band.}

The $H\alpha$ observations were taken with a different camera and were therefore registered with the $V$- and $I$-band images to have the same World Coordinate System astrometry. For this we used as a reference a bright, unsaturated source visible in all images.

Since the telescope point spread function extends beyond our adopted aperture radius of 2 pixels, it is necessary to correct  {the measured stellar fluxes} for the fraction of the total energy that falls outside this aperture. Following the encircled–energy curves for the ACS/WFC instrument derived by \cite{encircledenergies2016AJ....152...60B},  {we computed an aperture–correction factor equal to the encircled–energy fraction at 2 pixels, minus the difference between the fractions at 5 and 3 pixels. This subtraction accounts for the flux contained in the background annulus (3–5 pixels) used in our photometric routine.}  {The resulting factor was applied to all instrumental fluxes to recover the total source flux.} As for the photometric zero points,  {these were taken from the ACS Zeropoint Calculator\footnote{\url{https://acszeropoints.stsci.edu/}} and from \cite{zeropointWFC32009wfc..rept...30K} for the WFC3 $H\alpha$ filter, and were used to convert the aperture–corrected fluxes into calibrated magnitudes according to the standard HST VEGAmag photometric system.} These values are listed in Table~\ref{tab:photometry}.

\begin{table}[h]
\caption{Encircled energies and zero-point magnitudes for the different bands.}
\begin{tabular}{lcc}
\hline
\hline
{Band}      & {Encircled energy fraction} & {Zero-point magnitude} \\ 
\hline
 $I$         & $0.528$                     & $25.520$                       \\
 $V$         & $0.594$                     & $25.733$                       \\
 $H\alpha$ & $0.469$                     & $19.920$                       \\
\hline
\end{tabular}
\label{tab:photometry}
\end{table}

We limited our study to the 25482 sources with a combined uncertainty in the $V$ and $I$ bands $\sigma_{\rm VI} \leq 0.1$. Considering the uncertainties in the H$\alpha$ image, which is noisier due to the narrow-band filter, we required $\sigma_{\rm H\alpha} \leq 0.3$ in addition to our $\sigma_{\rm VI}$ uncertainty selection.  {This further constraint limits to 12649 the total number of sources that we consider to have solid photometry}. Table \ref{tab:cds} lists the position, magnitude, and uncertainty for the first few sources. The full table is available online.  

\begin{table*}
           
\label{tab:cds}      
\centering          
\caption{Photometric catalogue for the 12649 sources with good photometry (extract).} 
\begin{tabular}{c c c c l l l l l} 
\hline\hline       
ID &         RA &        Dec &    $V$ &  $\delta V$ &    $I$ &  $\delta I$ &    $H\alpha$ &   $\delta H\alpha$ \\
\hline                    
1 & 14.661975 & -72.357890 & 27.62 & 0.08 & 23.97 & 0.03 & 24.42 & 0.23 \\
2 & 14.659048 & -72.357840 & 25.35 & 0.03 & 19.88 & 0.01 & 20.77 & 0.04 \\
3 & 14.662388 & -72.357700 & 23.49 & 0.01 & 22.64 & 0.02 & 22.29 & 0.09 \\
4 & 14.657059 & -72.357651 & 23.49 & 0.01 & 22.62 & 0.02 & 23.13 & 0.13 \\
5 & 14.652981 & -72.357649 & 24.73 & 0.02 & 23.70 & 0.03 & 23.28 & 0.14 \\
6 & 14.660612 & -72.357628 & 24.40 & 0.02 & 23.47 & 0.03 & 24.24 & 0.21 \\
7 & 14.651501 & -72.357611 & 23.32 & 0.01 & 22.42 & 0.02 & 22.94 & 0.12 \\
8 & 14.664865 & -72.357595 & 23.93 & 0.01 & 23.01 & 0.02 & 22.62 & 0.10 \\
9 & 14.662593 & -72.357611 & 24.72 & 0.02 & 23.67 & 0.03 & 23.93 & 0.18 \\
10 & 14.652131 & -72.357592 & 23.44 & 0.01 & 22.58 & 0.02 & 22.83 & 0.11 \\
\hline                  
\end{tabular}
\tablefoot{For each source, the table lists the identification number (ID), the coordinates (RA and Dec in decimal degree), and the magnitudes in the $V$, $I$, and $H\alpha$ bands, together with their uncertainties. Only the first ten lines are shown here; the complete catalogue is available in electronic form at the CDS.}
\end{table*}

With the exposure times used in the $V$ and $I$ bands (see Table\,\ref{tab:imaging_data}), the  brightest objects in the field are saturated or affected by photometric non-linearity. To be conservative, we considered all stars brighter than $V = 19$\, {mag} to be affected by lack of linearity or saturation and retrieved their magnitudes using \textit{Gaia}. For each of these sources, we visually matched it to the corresponding object in the \textit{Gaia} Data Release 3 (DR3) catalogue \citep{gaiadr3} and retrieved the respective $G, G_{BP}$ and $G_{RP}$ magnitudes. These were converted to $V$ and $I$ magnitudes using known photometric conversions\footnote{\url{https://gea.esac.esa.int/archive/documentation/GEDR3/Data_processing/chap_cu5pho/cu5pho_sec_photSystem/cu5pho_ssec_photRelations.html#Ch5.T7}}. Of the original 167 sources with $V < 19$  {mag}, 100 have their magnitudes retrieved from the \textit{Gaia} catalogue. 

\begin{figure}
   \centering
   \includegraphics[width=9.5cm]{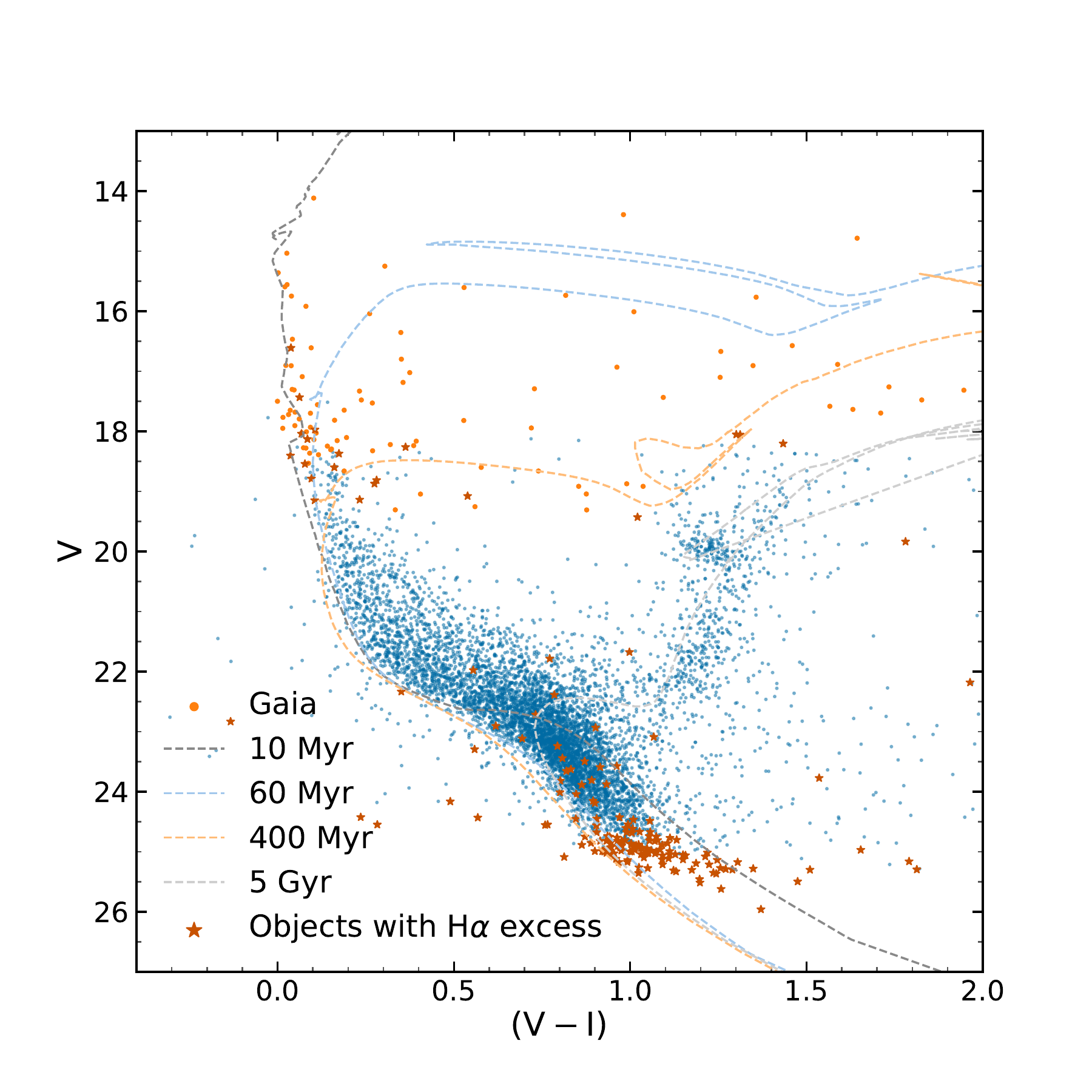}
     \caption{CMD showing the $V$ magnitude as a function of the  $V-I$ colour of the stars in the studied region. Shown in orange are the (potentially) saturated stars that were recovered through \textit{Gaia}. In brown are the objects with H$\alpha$ excess emission as identified in Section \ref{sec:pms}. The dashed lines represent isochrones for 10 Myr, 60 Myr, 400 Myr, and 5 Gyr.}
         \label{fig:corrcmd}
   \end{figure}

The resulting CMD is shown in Fig. \ref{fig:corrcmd}.  Clearly visible is the MS, as well as a red-giants red clump around $V \simeq 19$\, {mag} and $(V-I) \simeq 1.0$\, {mag}, with the majority of stars populating the bottom of the MS. The objects with magnitudes retrieved from \textit{Gaia} are indicated as orange dots in Fig. \ref{fig:corrcmd}.

{Also shown in the CMD are isochrones obtained from } \cite{padovatracks2012MNRAS.427..127B}, \cite{padovaextsmc2019MNRAS.485.5666P}, \cite{padovalowz2014MNRAS.445.4287T}, and \cite{padovaphotometric2019A&A...632A.105C}, for the specific HST ACS/WFC photometric system \citep{padovahstacs2008PASP..120..583G} and metallicity $Z = 0.004$. The isochrones correspond to ages of 10, 60, 400 Myr, and 5 Gyr.  Although the slightly elongated shape of the red clump suggests that some differential reddening might be present in the field \citep[see e.g.][]{extinction30dor2014MNRAS.445...93D}, its effects are small and for the reddening correction we only consider the component due to the Galactic foreground, corresponding to $E(V-I) = 0.16$\, {mag}, following the approach of \cite{starformationhistorySMCandNGC3462007AJ....133...44S},  {with a spread of about $0.04$\,mag \citep{hennekemperetal2008}}. Indeed, after correcting the isochrones for this foreground extinction, they follow the observations closely. We see that the brightest stars reach $V \simeq 15$\, {mag} and closely follow the 10 Myr isochrone, corresponding to a young population of recently formed massive stars. At $V \simeq 16$\, {mag}, we see signs of a population of roughly 60 Myr old red supergiants. Further down, there is a clear horizontal group, which we associate with an  {old} population of around 400 Myr old. Finally, the oldest stars in the field correspond to an age of several gigayears.

\section{Pre-main-sequence stars}\label{sec:pms}

Using our photometric catalogue, we searched for PMS candidates by identifying stars with significant H$\alpha$ emission.

\begin{figure}
   \centering
   \includegraphics[width=9.5cm]{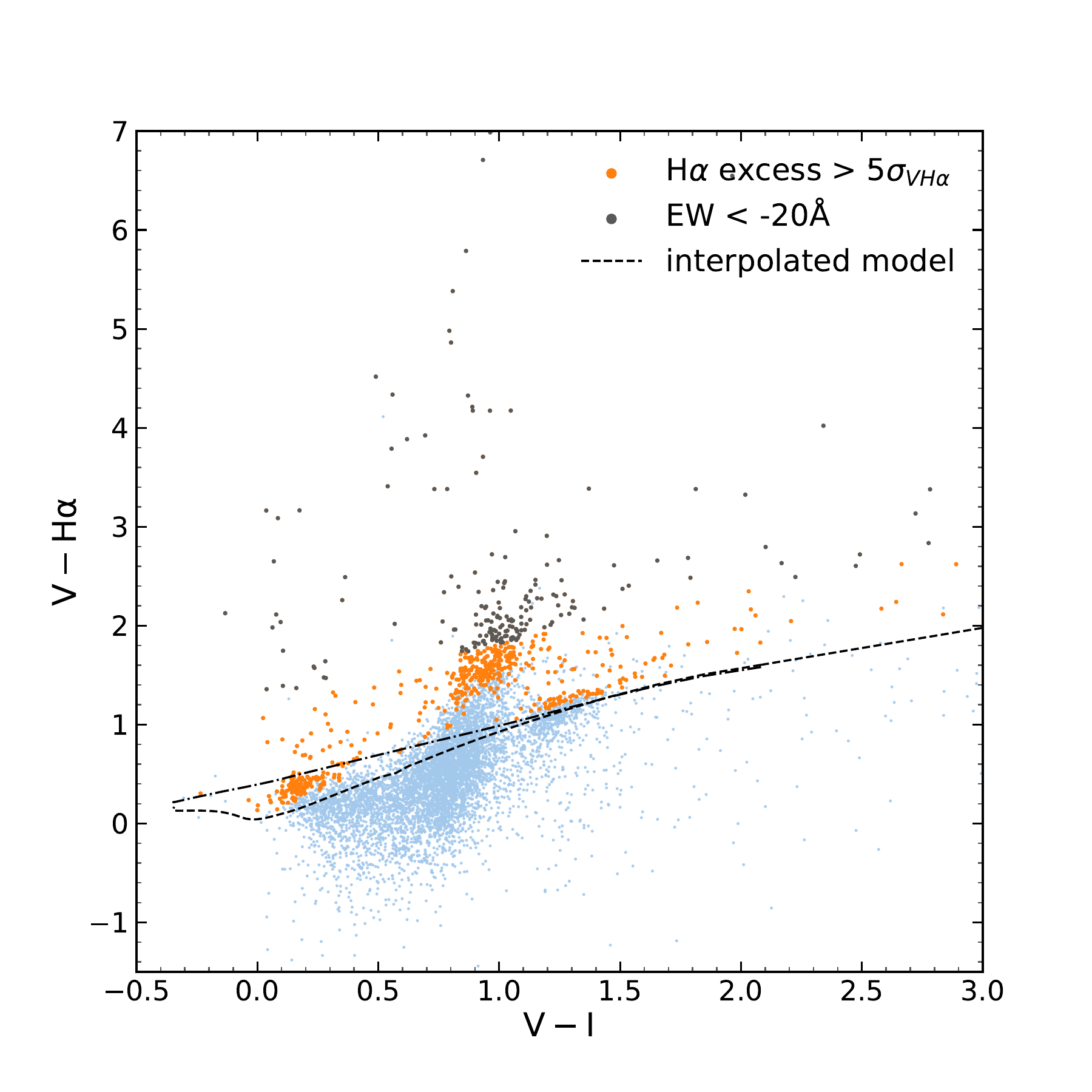}
     \caption{Colour-colour diagram of the sources in the studied region with $\delta_3 \leq 0.08$ {mag}. The dashed line indicates the curve interpolated from the model  {atmospheres} by \cite{bessel1998A&A...333..231B}, whereas the dot-dashed line indicates the continuum.  In orange are the objects with H$\alpha$ excess greater than 5 times the  {uncertainty on} $V-H\alpha$. The dark grey points are the sources that exhibit an emission equivalent $|W_{\rm eq}(H\alpha)| \geq 20$\,\AA.}
         \label{fig:pms}
   \end{figure}

{Pre-main-sequence candidates were identified through their H$\alpha$ excess in the $(V-H\alpha)$ versus $(V-I)$ colour--colour diagram (Fig.\,\ref{fig:pms}).} In particular, we searched for sources whose $V-H\alpha$ colour exceeds that of `normal’  { (i.e. MS) stars with the same $V-I$ by at least $5\,\sigma_{\rm VH\alpha}$ or $5$ times the combined photometric uncertainty in $V$ and $H\alpha$ (hereafter $5\,\sigma$ excess)}. We used model atmospheres by \citet{bessel1998A&A...333..231B} for the specific metallicity of the SMC ( {$1/8 < Z/Z_\odot < 1/5$}; \citealt{russelldopita1992, rollestonetal1999}) to determine the expected $V-H\alpha$ colour of  {MS stars (dashed line in Fig.\,\ref{fig:pms}}).  {Stars whose observed $V-H\alpha$ colour exceeds the photospheric value by at least $5\sigma$ were flagged as H$\alpha$-excess sources (orange dots in Fig.\,\ref{fig:pms}).}

{We then estimated the H$\alpha$ equivalent width $W_{\rm eq}(H\alpha)$ from the observed H$\alpha$ magnitude}, following the photometric method discussed in \citet{SN1987A2010ApJ...715....1D}.  {This allows us to quantify the strength of the emission line even without spectroscopy.} Besides requiring that objects have H$\alpha$ excess  {$>5\sigma$}, we also imposed the condition that the H$\alpha$ emission equivalent width must be at least $20$\,\AA\ in absolute terms, in order to exclude low-mass stars possibly affected by chromospheric activity \citep[see e.g.][]{EW2003ApJ...582.1109W}.  {Finally, we imposed a photometric quality criterion requiring that the combined uncertainty,}
\[
\delta_3 = \sqrt{\frac{\sigma_{\rm V}^2 + \sigma_{\rm I}^2 + \sigma_{\rm H\alpha}^2}{3}},
\]
 {be $\leq 0.08$ {mag}, as in \citet{SN1987A2010ApJ...715....1D}, where $\sigma_{\rm V}$, $\sigma_{\rm I}$, and $\sigma_{\rm H\alpha}$ are the photometric uncertainties in the three bands. Only stars meeting all three conditions ($\geq 5\sigma_{VH\alpha}$ excess, $|W_{\rm eq}(H\alpha)| \geq 20$\,\AA, and $\delta_3 \leq 0.08$ {mag}) were retained as robust PMS candidates with significant H$\alpha$ emission.}

Of the 12649 objects with  {solid photometry selected above}, 7302 sources satisfy this combined uncertainty constraint, and are plotted in dark grey in Fig.\,\ref{fig:pms}. Out of this sample, a total of 192 sources have both a 5$\sigma$ significant H$\alpha$ excess and $|W_{eq}(H\alpha)| \geq 20 \mathrm{\AA}$  {in emission}, and can therefore be identified as bona fide PMS stars.

We note that the sample might still include Be stars, which usually exceed the set equivalent width threshold as a result of strong stellar winds \citep[see e.g.][]{Bestars2005ApJ...622.1052M}. As we discuss in the next section, these stars will be excluded and our selection will be further narrowed down when we determine the physical properties of the candidate PMS objects.

The PMS candidates are shown in brown in the CMD in Fig. \ref{fig:corrcmd}. The majority of sources are fainter than $V\simeq 24$\, {mag} and there are objects both on the `red side' of the MS, where younger objects are expected, and closer to the MS itself, where more evolved PMS stars should be. At the top of the distribution, at magnitudes $V\lesssim 19.5$, we have another group of sources, which likely contain the aforementioned Be stars.

\subsection{Properties of PMS stars}
 {Following \cite{SN1987A2010ApJ...715....1D}, the} effective temperature ($T_{\rm eff}$) was obtained by interpolating the model atmospheres by \cite{bessel1998A&A...333..231B} in the specific HST bands of our study, as a function of the $V-I$ colour. Using the same model atmospheres, we determined the radius of the star in solar units as
\begin{equation}
    R = \sqrt{10^{-0.4(M_{\rm V} - M_{\rm V, ref})}}
,\end{equation}
where $M_{\rm V}$ is the absolute $V$ magnitude, calculated using a distance modulus of $18.98$ from the distance to the SMC of $62.4$ kpc \citep{distanceSMC2020ApJ...904...13G}. $M_{\rm V, ref}$ is the absolute $V$ magnitude derived from the theoretical \cite{bessel1998A&A...333..231B}  {model atmospheres} for a 1\,R$_{\odot}$ star. With the radii and effective temperatures of the stars, the bolometric luminosities ($L_{\rm bol}$) were derived, in solar units, from the Stefan--Boltzmann law, using  {a solar effective temperature} $T_\odot=5778$ K \citep{cox2000}. The typical uncertainty on $T_{\rm eff}$ is of the order of 150 K, while that on $L_{\rm bol}$ is approximately 5\,\%.

\begin{figure}
   \centering
   \includegraphics[width=9cm]{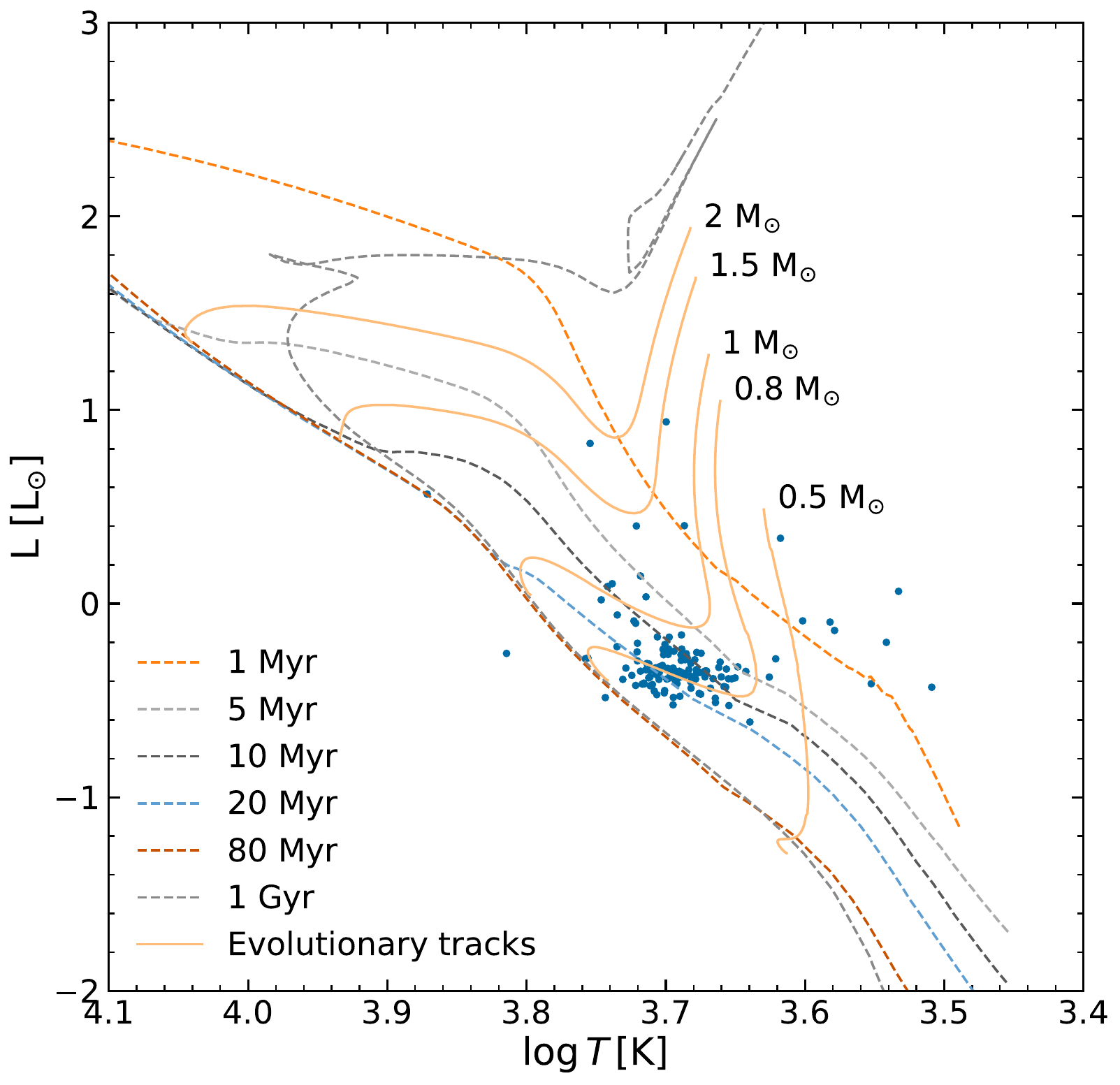}
     \caption{HR diagram of the PMS candidates in the studied region.  {Shown are the isochrones and evolutionary tracks from \cite{pisatracks2011A&A...533A.109T}}. The blue points indicate bona fide PMS stars.}
         \label{fig:hrd}
   \end{figure}

From the derived $L_{\rm bol}$ and $T_{\rm eff}$ , we  {constructed} a Hertzsprung--Russell (HR) diagram, as shown in Fig. \ref{fig:hrd}. As a reference, the evolutionary tracks  {of} \cite{pisatracks2011A&A...533A.109T} are added for masses in the range $0.5-2.0$ $\mathrm{M_{\odot}}$, ages younger than 100 Myr, and metallicity $Z=0.004$, as appropriate for the SMC.  We only include in this diagram the location of the 137 bona fide PMS stars, to which the evolutionary tracks apply. Only for these stars can a reliable mass and age be obtained (for the same reason, we have also excluded likely Be stars).

{Stellar masses and ages of the bona fide PMS candidates were derived using a probabilistic comparison with evolutionary tracks, following the `sieve' method outlined by \cite{sieve1998PhDT........21R} and later refined by De Marchi et al. (\citeyear{PMSngc346guido2011ApJ...740...11D}, \citeyear{guido30dor2011ApJ...739...27D}, \citeyear{2013ApJ...775...68D}, \citeyear{metallicity30dor2017ApJ...846..110D}). The PMS evolutionary tracks are first resampled to build a fine grid in the HR diagram, interpolated between major evolutionary phases. For each star, the observational uncertainties in $T_{\rm eff}$ and $L_{\rm bol}$ define a cell in the HR diagram. All evolutionary tracks crossing this cell contribute possible age--mass solutions, weighted by the evolutionary time spent in the cell. The resulting probability distributions yield the most likely mass and age and their associated uncertainties.} The procedure is discussed by \cite{metallicity30dor2017ApJ...846..110D}, to whom we refer the reader for further details. In the end, we were able to derive physical parameters for all 137 PMS stars and to measure the associated statistical uncertainties.

{While the positions of PMS stars in the HR diagram allow us to estimate their masses and ages, it is well known that individual PMS ages derived from isochrone comparison can be highly uncertain, owing to the tight spacing of isochrones at early evolutionary stages \citep[see e.g.][]{baraffeetal2002,hillenbrandwhite2004,belletal2013,jeffriesetal2017}. Masses are generally more robust, since evolutionary tracks of different masses remain well separated over the relevant temperature range. In the low-metallicity environment of the Magellanic Clouds, however, PMS stars evolve more rapidly across the HR diagram, and the isochrones are correspondingly more widely spaced, reducing the relative age uncertainty by more than a factor of 2 compared to solar metallicity regions \citep{demarchietal2024}.}

\begin{figure*}
\resizebox{\hsize}{!}
        {\includegraphics{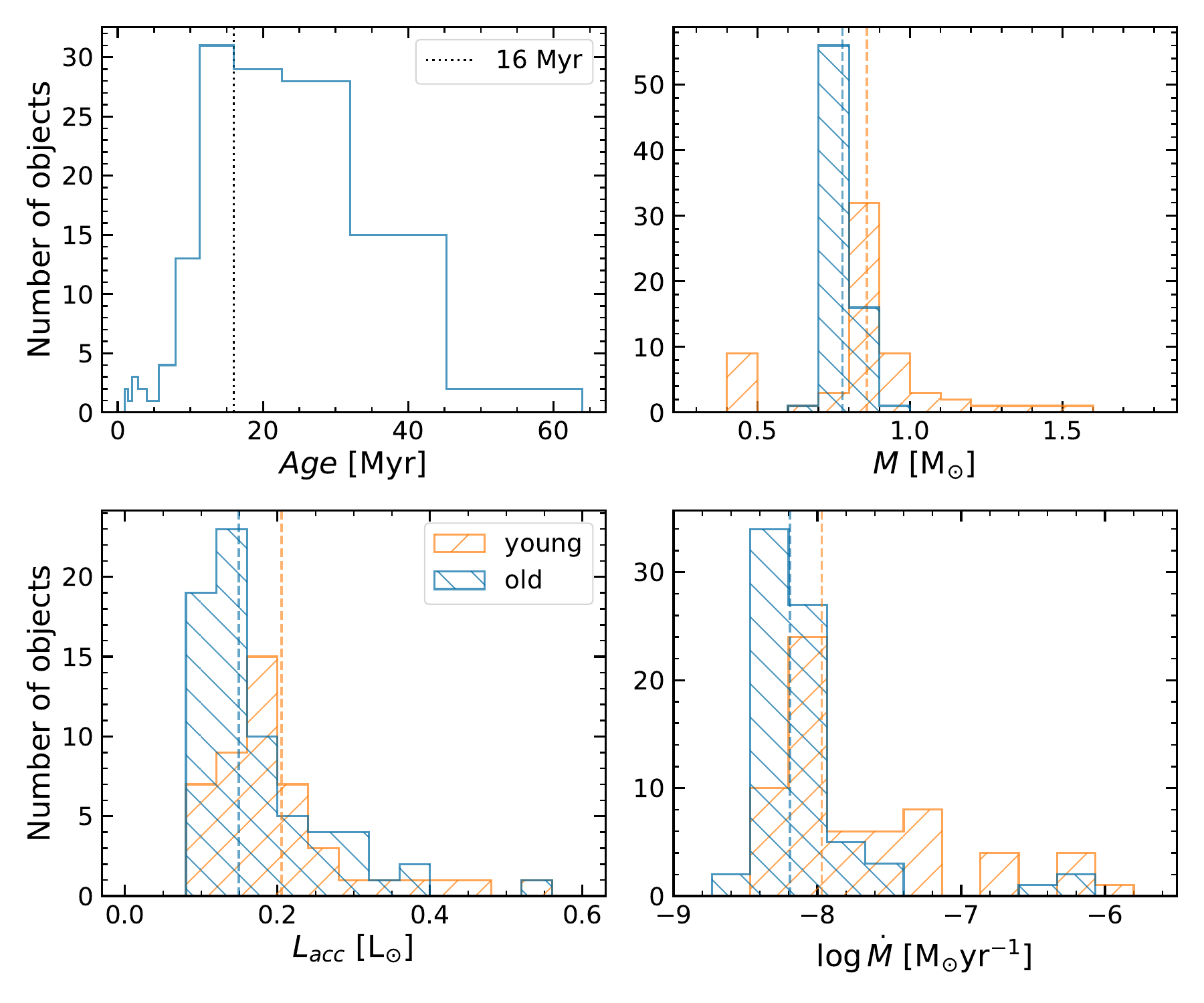}}
        \caption{Histograms showing the distribution of age (\textit{top left}), mass (\textit{top right}), accretion luminosity (\textit{bottom left}), and mass accretion rate (\textit{bottom right}) determined for the 137 bona fide PMS stars. The blue (orange) histograms represent the stars older (younger) than 16 Myr. Dashed vertical lines indicate the median of the respective distribution.}
    \label{fig:hists}
\end{figure*}

The distribution of derived ages and masses is shown in the top  {panels} of Fig. \ref{fig:hists}. The uncertainties on $\log T_{\rm eff}$ and $\log L_{\rm bol}$ primarily come from the photometry and are roughly constant across the CMD (and HR diagram), but as the age increases, the separation between isochrones becomes progressively smaller. A typical $V-I$ difference of $0.1$\,mag between the colours of two PMS stars in the CMD could imply an age difference of $\sim$2 Myr at ages around 4 Myr, growing to a difference of $\sim$7 Myr around at an age of 20 Myr. For this reason, in Fig. \ref{fig:hists} we adopted for the ages a logarithmic step of $\sqrt{2}$, which is always wider than the observational uncertainties (see \citealt{guido30dor2011ApJ...739...27D}).  {We used the median age of 16 Myr to split the population in a `young' and an `old' group, containing respectively 63 stars (shown in orange in Fig. \ref{fig:hists}) and 74 stars  (blue).} 

Considering the masses of the PMS stars, there is a difference between the  {young} and  {old} populations.  {The top-right panel of Fig.\,\ref{fig:hists} shows the PMS mass distribution. Most stars are subsolar in mass, around $0.7-0.9$ $\mathrm{M_{\odot}}$. The { mass distribution of old PMS stars} is sharply peaked with a mean of $0.77$\,M$_\odot$ {and} a standard deviation of $0.05$\,M$_\odot$. The young PMS population has two peaks, with a small group of stars at $0.45$\,M$_\odot$ and a much more numerous and broader distribution peaked at $0.85$\,M$_\odot$ with a standard deviation of $0.22$ \,M$_\odot$. The median mass values are also shown in the figure (vertical dashed lines).}

{The observed pattern is not unexpected. At the earliest stages of star formation, stars of all masses are still in the PMS phase and typically show active accretion, and hence significant $H\alpha$ excess. As time goes by, however, accretion progressively weakens, and this decline occurs more rapidly in the more massive PMS stars, which contract and reach the MS on shorter timescales. As a consequence, at later evolutionary stages the sources that still satisfy our strict $|W_{\rm eq}(H\alpha)| > 20$\,\AA\ criterion are predominantly of lower mass, whereas the more massive objects have already ceased accretion and no longer meet this requirement. Obviously, many additional sources (both young and old) show weaker $H\alpha$ excess (orange points in Fig.~\ref{fig:pms}), but, as discussed in Section~4, their nature is less certain because they may include objects dominated by chromospheric activity rather than accretion.}

\subsection{Accretion luminosity and mass accretion rate}

{The H$\alpha$ luminosity was computed from the observed line flux at the stellar surface as} 
\begin{equation}
    L_{\rm H\alpha} = 4\pi \, R^2 \, F_{\rm H\alpha} ,
\end{equation}

{\noindent using the standard definition of luminosity for a spherical emitter \citep[see e.g.][]{mihalas1978}. In our specific case} $F_{\rm H\alpha}$ is the difference between the observed H$\alpha$ flux and the theoretical H$\alpha$ flux from the  {model atmospheres} by \cite{bessel1998A&A...333..231B}. In turn, the observed $F_{\rm H\alpha, obs}$ flux is obtained by multiplying the observed count rate in the $H\alpha$ band by the specific inverse sensitivity $\eta$ and by the rectangular width $RW$ of the F656N band, as per the equation

\begin{equation}
    F_{\rm H\alpha, obs} = 10^{-0.4(m_{\rm H\alpha}-m_{\rm 0})} \, \eta \,\, {RW}.
\end{equation}
The inverse sensitivity $\eta$ represents the flux that a source needs to have to produce a count rate of 1 e\,s$^{-1}$ for a constant $F_{\lambda}$ \citep{photflam2020acs..rept....8B} and corresponds to $1.538 \times 10^{-17}$ erg cm$^{-2}$ s$^{-1}$\,\AA~ for the specific H$\alpha$ filter of the HST/WFC3 camera. {For} the same instrumental set-up $RW = 17.68$\,\AA, {while} $m_0$ is the zero-point magnitude for H$\alpha$ (Table \ref{tab:photometry}). 
The H$\alpha$ flux is then obtained by simply subtracting the theoretical flux ($F_{\rm H\alpha, B}$) from the observed flux ($F_{\rm H\alpha, obs}$):
\begin{equation}
    F_{\rm H\alpha} = F_{\rm H\alpha, obs} - F_{\rm H\alpha, B}.
\end{equation}

\begin{table*}[h]
\label{tab:Mdotcomp}      
\centering          
\caption{Mass accretion rates for PMS stars inside NGC 346 proper and in the region studied here, per age group.}
\begin{tabular}{l|cc|cc}
\hline
\hline
               & \multicolumn{2}{c|}{{NGC 346 proper}}                                          & \multicolumn{2}{c}{{NGC 346 periphery (this work)}} \\
 {Age group} &  {Median $ \dot M_{\rm acc}$}  &  {No. of objects} &  {Median $ \dot M_{\rm acc}$}     &  {No. of objects}   \\ \hline
Younger than 8 Myr              & $(1.1 \pm 0.7) \times 10^{-7}$                                                 & 345                     & $(6 \pm 4) \times 10^{-8}$                        & 19                        \\
Between 8 and 20 Myr              & $(5.0 \pm 2.0) \times 10^{-8}$                                                 & ~~75                      & $(8 \pm 3) \times 10^{-9}$                       & 59                        \\
Older than 20 Myr              & $(1.5 \pm 0.6) \times 10^{-8}$                                                 & 412                     & $(6 \pm 2) \times 10^{-9}$                        & 59 \\                   \hline   
\end{tabular}
\tablefoot{For each age group, the median mass accretion rates and corresponding median absolute deviations (in units of M$_\odot$ yr$^{-1}$) are indicated, together with number of objects in that group.}
\end{table*}

The relationship between the accretion luminosity and the H$\alpha$ luminosity is unknown, but previous studies in this series
\citep{PMSngc346guido2011ApJ...740...11D, 2013ApJ...775...68D, metallicity30dor2017ApJ...846..110D, 2015A&A...574A..44B, 2019ApJ...875...51B, 2022A&A...663A..74C, tsiliaetal2023, vlasblomdemarchi2023} have followed the  {empirical relationship} between $L_{\rm H\alpha}$ and $L_{\rm acc}$ derived by \cite{SN1987A2010ApJ...715....1D}, based on the analysis of the observations of a number of T Tauri stars by \cite{Lacc2008AJ....136..521D}. That relationship is based on the assumption that $L_{\rm acc}$ and $L_{\rm H\alpha}$ are linearly related \cite[see e.g.][]{clarkepringle2006MNRAS.370L..10C},  {in a magnetospheric accretion scenario}.

More recent studies by \cite{LaccLhaprop2017A&A...600A..20A} of a larger sample of T Tauri stars in Lupus  {revisited} the matter and the authors  {concluded} that a purely empirical fit to the data suggests a non-linear relationship. However, in the range of interest in this work, $0.1 \la L(H\alpha) \la 0.6$, the differences in $L_{\rm acc}$ implied by the two relationships is considerably smaller than the uncertainties intrinsic to the relationships themselves. Therefore, for the sake of comparison with previous works by our team, we continue to use the linear relationship of \cite{SN1987A2010ApJ...715....1D} given by
\begin{equation}
    \log L_{\rm acc} = (1.72 \pm 0.25) + \log L_{\rm H\alpha} .
\label{eq:L_acc}
\end{equation}

{We note that the observations in the $H\alpha$ band are not simultaneous with those in the continuum bands (obtained with both HST and \textit{Gaia}). This can introduce an uncertainty on the derived H$\alpha$ luminosity. To estimate its impact, it is worth recalling that accretion‑dominated T Tauri stars exhibit optical colour changes of typically $\Delta(V-I) \simeq 0.1-0.4$\,mag, reaching $\simeq 0.5$\,mag in strong accretors, and they become systematically bluer when brighter \citep{codyetal2014,venutietal2015,staufferetal2014}. With reference to Fig.\,\ref{fig:pms}, this implies a difference of at most $\sim 0.2$\,mag in the $V-H\alpha$ colour for the continuum under H$\alpha$,  {indicated by the dot-dashed line}. Since our bona fide PMS candidates ({orange} dots in Fig.\,\ref{fig:pms}) have a median $\Delta(V-H\alpha) \simeq 1.2$\,mag, the resulting  uncertainty on $L(H\alpha)$ is about 20\,\%. In fact, in most cases $L(H\alpha)$ will be brighter by that amount, since we are more likely to detect the H$\alpha$ excess during a burst, when $(V-I)$ is bluer.}

The accretion luminosity derived with Eq. \ref{eq:L_acc} can be converted to a mass accretion rate ($\dot{M}_{\rm acc}$) through the free-fall equation
\begin{equation}
    L_{\rm acc} \simeq \frac{G M_{\ast} \dot{M}_{\rm acc}}{R_{\ast}} \left( 1 - \frac{R_{\ast}}{R_{in}} \right),
\end{equation}
{where $G$ is the gravitational constant, $M_{\ast}$ is the mass of the star, $R_{\ast}$ its} (photospheric) radius, and $R_{in}$ the inner radius of the accretion disc. The {value of} $R_{in}$ depends on factors such as the coupling of the accretion disc to the stellar magnetic field, which we cannot measure with our observations. Therefore, {we adopted} $R_{in} = 5 R_{\ast}$ for each of the PMS stars, following \cite{accretionradius1998ApJ...492..323G}. 


The resulting distribution of accretion luminosities and mass accretion rates can be found in the bottom panels of Fig. \ref{fig:hists}. There is no significant difference between the median accretion luminosity and mass accretion rate for the  {young} and  {old} PMS stars within the uncertainties,  {but}  {old} PMS objects  {tend to} have lower accretion luminosities and mass accretion rates, as one would expect in general. A median mass accretion rate of $ \dot M = (1.1 \pm 0.5) \times 10^{-8} \mathrm{M_{\odot} yr^{-1}}$ and $(6.5 \pm 1.7) \times 10^{-9} \mathrm{M_{\odot} yr^{-1}}$, with  {the uncertainties being the median absolute deviation}, are found for the  {young} and  {old} group, respectively. These values are lower than those found with the same method for the main star forming cluster NGC 346 \citep{PMSngc346guido2011ApJ...740...11D}, and recently confirmed spectroscopically by \cite{demarchietal2024}, although in our sample there are a number of young PMS objects with comparable mass accretion rates between $10^{-8}$ and $10^{-7} \mathrm{M_{\odot} yr^{-1}}$.


To  {further}  {investigate} the difference in $\dot M_{\rm acc}$  {between the outer field studied here and NGC~346 proper (see Fig.\,\ref{fig:fields})}, we took the $\dot M_{\rm acc}$ values of all PMS stars in the main cluster from \cite{PMSngc346guido2011ApJ...740...11D} and  {divided both that sample and our} PMS sample into three age groups.  {The ages were chosen so that the sample in our outer-field is split into roughly equal-number subsets, namely} objects younger than 8~Myr, older than 20~Myr, and those in between.  {The median mass accretion rates of the three groups and their uncertainties are} shown in Table~4. Clearly, $\dot M_{\rm acc}$ is consistently lower for stars of the same age and mass in the outer field studied here  {than in the central NGC~346 cluster},  {as has already been observed for other lower density star-forming regions in the Magellanic Clouds} \citep[see e.g.][]{2019ApJ...875...51B, 2022A&A...663A..74C, vlasblomdemarchi2023, tsiliaetal2023}.

\section{Effect of nearby  {massive} stars}\label{sec:nearby_stars}

\begin{figure}[ht]
    \resizebox{\hsize}{!}
        {\includegraphics{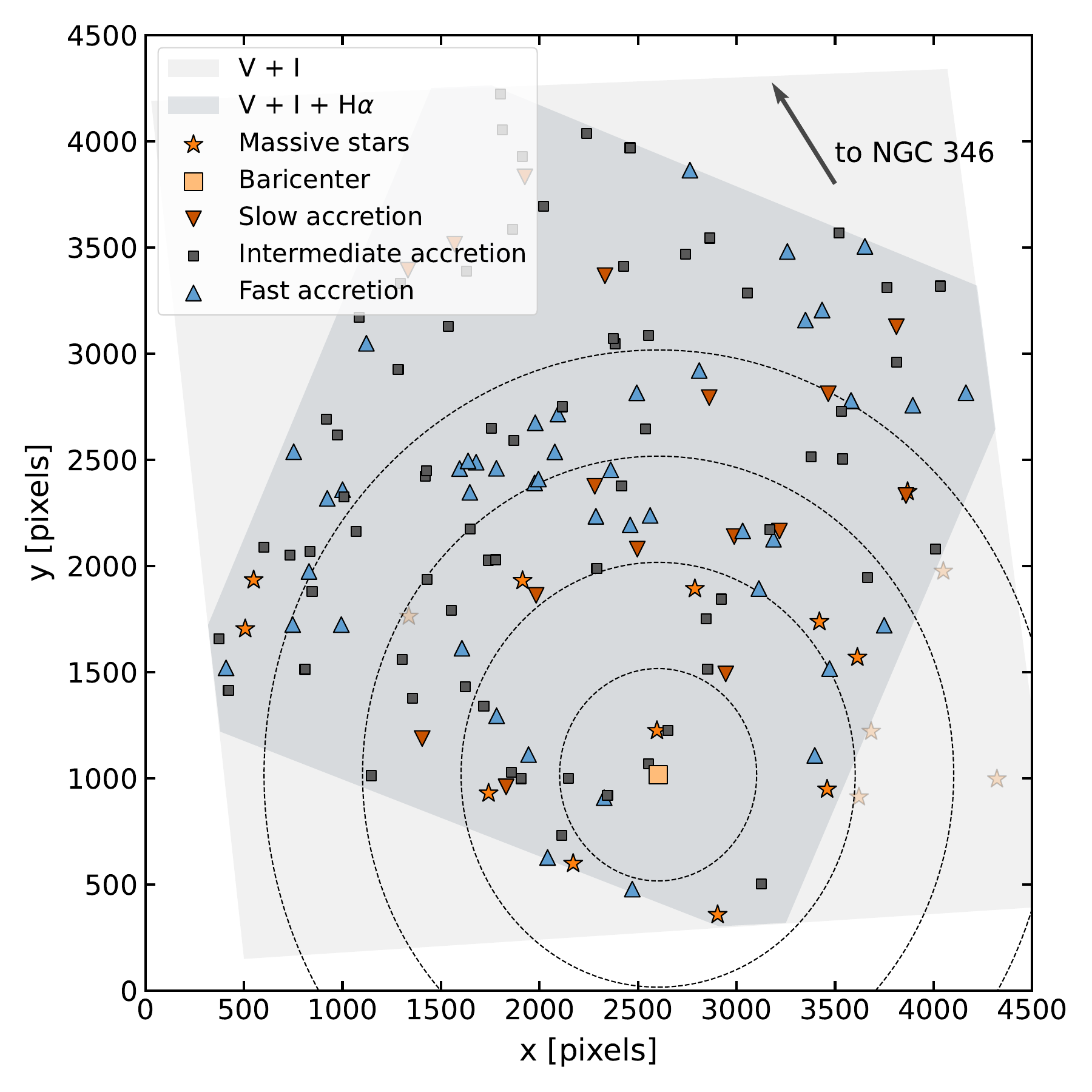}}
    \caption{Distribution of PMS stars in the field, colour-coded by their relative mass accretion rate: in blue  `fast' accretors ($\dot{M}_{\rm acc} > 10^{-8} \mathrm{M_{\odot} yr^{-1}}$), in dark grey `intermediate' accretors ($10^{-8.3} \mathrm{M_{\odot} yr^{-1}} < \dot{M}_{\rm acc} < 10^{-8}\, \mathrm{M_{\odot} yr^{-1}}$), and in red `slow' accretors ($\dot{M}_{\rm acc} < 10^{-8.3} \mathrm{M_{\odot} yr^{-1}}$).  The most massive young (non-PMS) stars are plotted in orange, with their barycentre in yellow. Radial shells around this barycentre are shown, as well as the outlines of the imaged fields {(see Fig.\,\ref{fig:fields})} and an arrow pointing towards the centre of the NGC 346 cluster proper.}
    \label{fig:barycentre}
\end{figure}

Given the proximity of the studied region to the  NGC 346 cluster itself, it is interesting to investigate whether there is a relationship between the location and physical properties of the PMS stars in this field and the massive stars in NGC 346. The relative distribution is shown in Fig.\,\ref{fig:barycentre} and we used it to look for any gradient in the direction of the cluster. As for the spatial distribution of the PMS stars, these objects are distributed rather uniformly across the field and we do not observe more active star formation in the direction of NGC 346, nor does the $L(H\alpha)$ luminosity of the PMS stars appear to correlate with their projected distance from NGC 346. This is not surprising, considering that the projected distance to the centre of NGC 346 is about 164 pc. 

However, the data reveal an interesting trend between the $L(H\alpha)$ luminosity of the PMS stars and their distance from the barycentre of a small group of massive stars in the west-southwest portion of the field, whose ionising radiation might be contributing to the dissipation of the circumstellar discs through photoevaporation \citep{2025A&A...701A.139R}. Following \cite{ SN1987A2010ApJ...715....1D}, we selected young massive stars with $\log (T_{\rm eff}/{\rm K}) \ge 4.4$  and $\log(L/L_\odot) \ge 4$, having adopted a distance of $62.5$ kpc \citep{distanceSMC2020ApJ...904...13G} and $A_{\rm V}=0.84$ (see Section\,\ref{sec:photometry}). We note that the values of $T_{\rm eff}$ were derived from the $V-I$ colours, using the model atmospheres of \cite{bessel1998A&A...333..231B}, and as such they might underestimate the actual effective temperatures. The 17 objects selected in this way, with $25,000$\,K$\, \la T \la 32,000$\,K, correspond to  {sources in the upper part of the MS} with $V<17.5$  {mag} and  $V-I<0.12$  {mag}.  {Their positions are shown by orange star symbols in Fig.\,\ref{fig:barycentre}.}  {Stars with a redder $V-I$ colour are not considered here, since we are interested in the effect of strong ultraviolet radiation originating from these massive objects}. In total, 12 of these sources fall within the field of the H$\alpha$ observations and have reliable H$\alpha$ photometry. Using the 10\,Myr isochrone in Fig. \ref{fig:corrcmd}, we derived approximate masses for these objects, all of which are larger than $\sim 10.5$\,M$_\odot$. We also used the 10 Myr isochrone to assign masses to stars in the range $17.5\,\mathrm{mag} < V < 18.5\,\mathrm{mag}$ and with $V-I<0.08$  {mag}, which we used to compute the mass function (see below) {. Since these sources have $T_{\rm eff} < 25,000$\,K, their ionising photon rate would not contribute significantly to the photoevaporation of the discs. }

The yellow square  {in Fig.\,\ref{fig:barycentre}} corresponds to the barycentre of {the} ionising photon rates {of the 17 massive stars,} taken for each source from \cite{Panagia1973AJ.....78..929P} {and} computed under the simplifying assumption that all objects are at the same distance from us on the plane of the sky. 
We note that all of the most massive stars are concentrated in the lower (southern) part of the image, suggesting a recent star formation episode in that part of the region.

\begin{figure}[t]
   \centering
   \includegraphics[width=9cm]{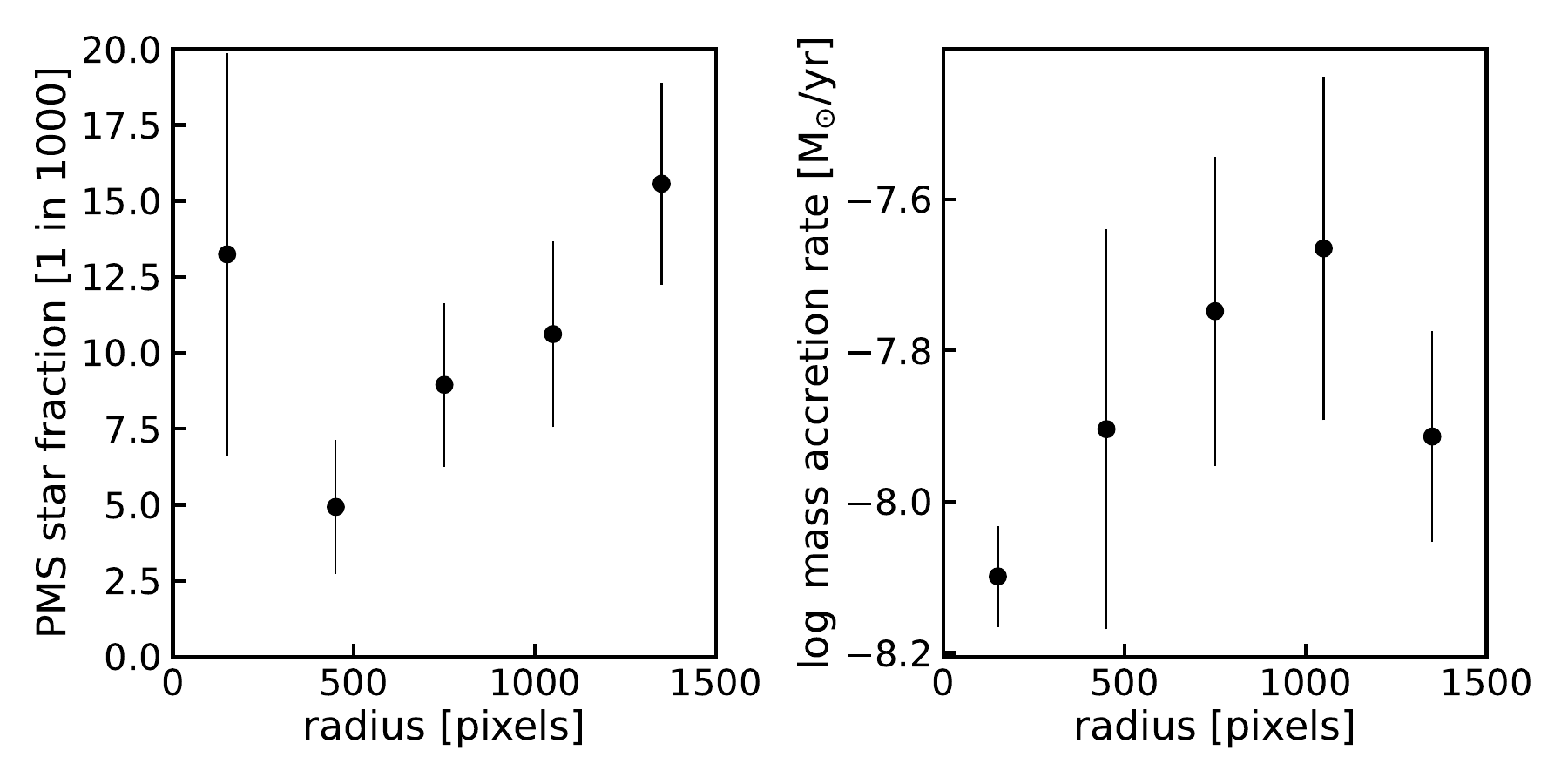}
     \caption{\textit{Left}: Fraction of PMS stars  {relative} to the total number of stars. \textit{Right}: $\dot M_{\rm acc}$ of the PMS stars averaged over each shell.}
         \label{fig:massive}
   \end{figure}

From the barycentre, we  {moved} radially outwards, dividing the region into concentric circular annuli (`shells') and  {considered} the stars in each consecutive shell,  {as indicated by dashed circular lines} in Fig. \ref{fig:barycentre}.  {The radial width of the shells is $15^{\prime\prime}$, corresponding to a projected distance of roughly $4.4$\,pc. The total number of bona fide PMS stars inside each shell  {was} counted, as well as the fraction of the total number of stars that they represent.} These fractions are shown in Fig. \ref{fig:massive} (left).  {In addition, the average mass accretion rate of the stars in each shell  {was} computed, as shown in the same figure on the right. The uncertainties shown in the figure represent the standard deviation within the shells.}  {As mentioned above, this is a simplifying two-dimensional projection, because the distance to each individual star is not known. However, we are considering an ordinary region of the SMC field, which is known to be dominated by old and intermediate‑age stars \citep[see e.g.][]{hodge1985, harriszaritsky2004}, so there is no reason to expect large anisotropies.}

{Indeed, the non-PMS stars in this field are distributed evenly in number across the region, so if also the PMS stars had a uniform distribution then one would expect the ratio of PMS and non-PMS stars to remain constant across the field. This would correspond to a horizontal trend in Fig. \ref{fig:massive}}. Similarly, if there were no influence due to the environment, the average $\dot M_{\rm acc}$ would remain constant across each shell. Instead, what is observed is a general increase in both the PMS star fraction and the mass accretion rate when moving radially away from the barycentre of the young massive stars.  {The uncertainties are largest in the central bin, where only six PMS stars are present, and the resulting fraction is affected by small‑number statistics}.  {(We note that the trends are still evident if the size of the shells is reduced to $10^{\prime\prime}$, albeit with a larger scatter.)}

{The observed trends suggest} that there  {might}  {be} an interaction between this young population of massive stars and the PMS stars.  {In particular, the radiation of the nearby massive stars might be eroding the discs, possibly through both ionisation and photoevaporation \citep[see e.g.][]{stoerzerhollenbach1999, adamsetal2004}, thereby reducing disc lifetimes and causing the accretion process to fade.}

{Of course, these trends must be interpreted with caution, since neither the PMS stars nor the massive stars responsible for the ionising radiation have known positions along the line of sight. The massive stars themselves are spatially scattered, and the lack of three-dimensional information introduces additional uncertainty in the location of their barycentre.}  {For this reason, while the data indicate the presence of an environmental effect possibly caused by the massive stars, the available information does not allow a fully quantitative characterisation of its strength.}

{To investigate the nature of these massive objects, it is useful to understand whether they formed in situ in their current location or are instead in some way related to NGC\,346. This helps in clarifying the role that NGC\,346 may have played in shaping the stellar populations observed in the surrounding region. To this end, we} considered the kinematics of the field. Using proper motions from the \textit{Gaia}  {DR3} catalogue \citep{gaiadr3} for the young massive stars, we find that the bulk motion is dominated by the proper motion of the SMC itself. In our sample, the median proper motions  {components towards the west and north} as measured by \textit{Gaia} are, respectively, $\mu_W = -0.75$ mas yr$^{-1}$ and $\mu_N = -1.27$ mas yr$^{-1}$  {(with sample standard deviations of $\sigma_{\rm \mu_W} = 0.21$ mas yr$^{-1}$ and $\sigma_{\rm \mu_N} = 0.12$ mas yr$^{-1}$)}, compared to the systemic proper motion  {components} of the SMC of $\mu_W = -0.82$ mas yr$^{-1}$ and $\mu_N = -1.21$ mas yr$^{-1}$. After subtracting  {the systemic components}, there remains no preferential direction of motion, nor  {evidence} of rotation,  {for any of the massive stars in our field. This suggests that the young massive stars have likely formed there, and there is no detectable effect of outflow or rotation due to the massive nearby NGC\,346 cluster.}

\begin{figure}[ht]
    \resizebox{0.9\hsize}{!}
        {\includegraphics{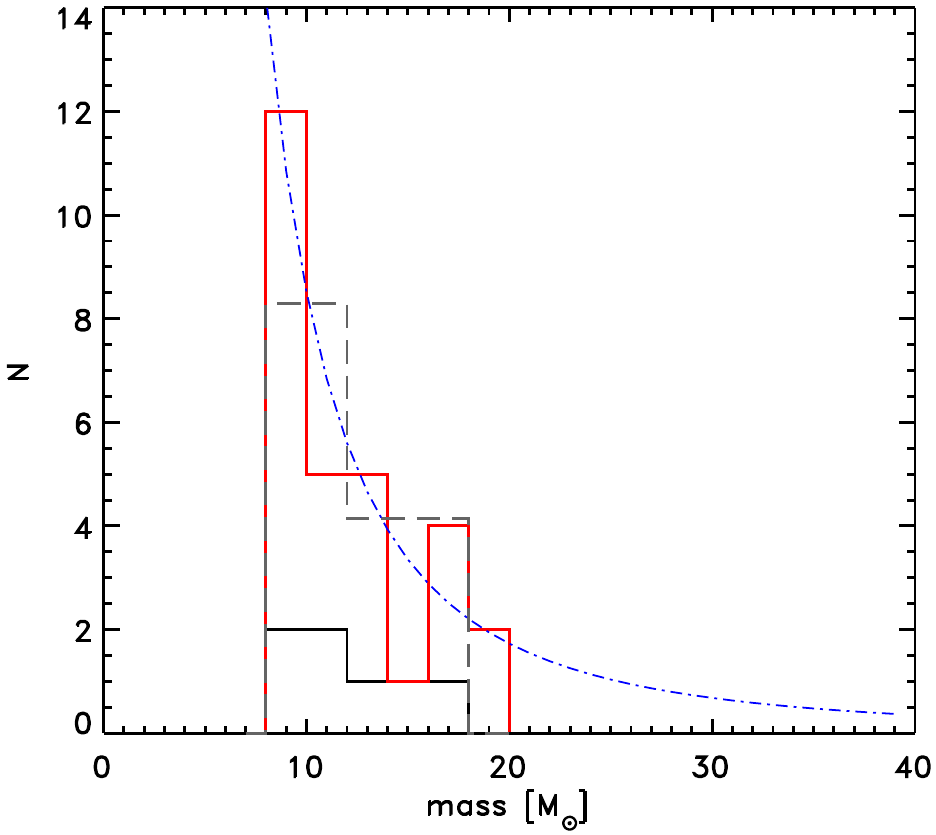}}
    \caption{Histograms of the number of massive stars ($8-20$\,M$_\odot$) in our region (red line) and in the ONC (black line). The dashed line represents the IMF of the ONC rescaled to match the number of objects under the red line. The best fitting standard \cite{Kroupa2002Sci...295...82K} IMF is shown by the dot-dashed line. }
    \label{fig:IMF}
\end{figure}

The area that we surveyed represents a good example of massive star formation in the field. It is known that a non-negligible fraction of massive stars in the Magellanic Clouds are found outside prominent clusters: while many are runaways \citep[see e.g.][]{Evans2011A&A...530A.108E},  {about} 35\,\% are consistent with in situ formation in sparse groups or isolation \citep{Oey2004AJ....127.1632O, Oey2013ApJ...768...66O}. To characterise the population of massive stars in this region, we computed the mass function using the masses estimated through comparison with the 10\,Myr isochrone (see above). {The distribution of stars with masses of $8 - 20$\,M$_\odot$ (red histogram in Fig.\,\ref{fig:IMF}) agrees well with the \cite{Kroupa2002Sci...295...82K} initial mass function (IMF; dot-dashed line). For comparison, we also consider the IMF of the Orion Nebula Cluster \citep[ONC;][]{Hillenbrand1997AJ....113.1733H, DaRio2010ApJ...722.1092D} over the same mass range (black solid line), together with a version rescaled by a factor of $4.1$ (dashed line) to match the number of objects observed in our region in the $8 - 20$\,M$_\odot$ interval.} The comparison is necessarily approximate because we derived the masses of our stars from their photometry and cannot assess their multiplicity, which is expected to be high for massive objects in general and even more so at low metallicities \citep{Sana2012Sci...337..444S, Shenar2024A&A...690A.289S}. The ONC currently hosts an even more massive object, the 34\,M$_\odot$ $\theta^1$ Ori C1 \citep{Kraus2007A&A...466..649K}, which is not included in the IMF of Fig.\,\ref{fig:IMF}  {because} in a field like ours, with an age of 10\,Myr and older, objects of that mass are expected to have already exploded as Type II supernovae  {and would no longer contribute to the mass function}.

The surface density of massive stars in our field is considerably lower than in a typical stellar cluster: the $8-20$\,M$_\odot$ stars are distributed over an area of $\sim 12$\,pc radius, whereas in the ONC they are contained within $\sim 0.6$\,pc. However, despite the lower surface density, the massive stars that we detect are more than four times those in the ONC in the same range. This comparison serves to highlight the potentially significant contribution of field star formation to the overall mass budget of galaxies.  

\section{Comparison with JWST observations}\label{sec:comparison}

\begin{figure}[ht]
   \centering
   \includegraphics[width=9cm]{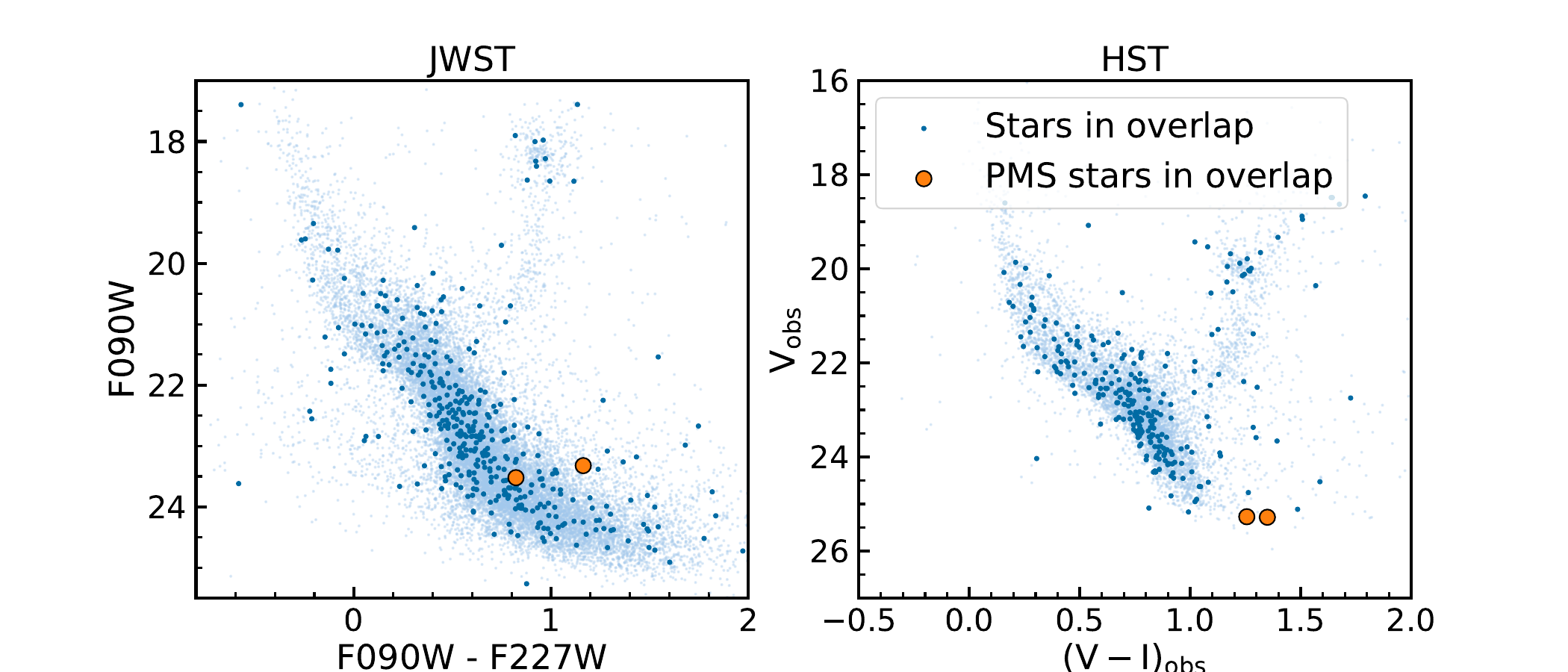}
     \caption{ {CMDs obtained with the JWST/NIRCam (\textit{left}) and HST (\textit{right}) photometry. In the left panel, light blue dots are JWST sources outside the HST--JWST overlapping region, and blue dots are sources only included in the field observed with JWST. In the right panel, blue marks the HST sources in the overlapping region. The orange points indicate the two bona fide PMS stars with reliable NIRCam photometry; they are shown in both panels.}}
         \label{fig:jwstcmd}
   \end{figure}

{JWST/NIRCam observations exist in a field partly overlapping with the one studied in this work (Fig.\,\ref{fig:fields}; see Section\,2). Because the NIRCam observations cover near-IR wavelengths ($0.9$--$4.3\,\mu$m), they are complementary in wavelength to our HST data, which probe the optical regime including H$\alpha$; together they allow us to detect both accretion signatures and circumstellar-disc emission from the same stars. In this work we used the photometric analysis of the NIRCam data provided by K. Fahrion (private comm.), following the workflow discussed in \cite{fahriondemarchi2024}, to identify and measure the magnitudes of the sources in the field observed with NIRCam around NGC 346 (see Fig. 1).  We  {constructed} a CMD in the F090W and F227W bands for the overlapping area,  {which is} shown in the left panel of Fig.\,\ref{fig:jwstcmd}. The right panel reproduces the HST CMD from Fig.\,\ref{fig:corrcmd} for the same region. }

{Of the 137 bona fide PMS stars, four fall within the field covered by NIRCam. Two of these lie too close to the detector edges and as such lack reliable photometry, leaving two well-measured PMS candidates for comparison. Their identification numbers are 12510 and 12635 in Table \ref{tab:photometry}.} One of the two PMS stars ends up with a significantly redder F090W-F227W colour than the bulk of MS stars, further strengthening the notion that this object is indeed a PMS star surrounded by a dusty circumstellar disc that causes some near-IR excess. Additionally, the other star, with a F090W--F227W colour of $0.8$, is consistent with the CMD position of PMS stars on the central NGC\,346 cluster as observed by \cite{2024ApJ...971..108H} with NIRCam.

The JWST photometry is complementary to the HST photometry and reveals different properties of the stars. While H$\alpha$ excess indicates accretion from the circumstellar disc onto the star, near-IR excess originates from thermal disc emission \citep[see e.g.][]{HavsnearIR2011MNRAS.410..227C}. Detecting both would be a strong sign of the PMS nature, although the absence of significant near-IR excess cannot rule out the PMS nature of the objects, particularly for  {old} PMS stars with evolved and transitional discs with inner dust cavities \citep[see e.g.][]{2010ApJ...710..597S}. Similarly lack of $H\alpha$ excess emission at the $5\,\sigma$ level of more does not imply that the star is not accreting, rather that accretion might be too weak to be detected with our method.

\begin{figure}[ht]
   \centering
   \includegraphics[width=9cm]{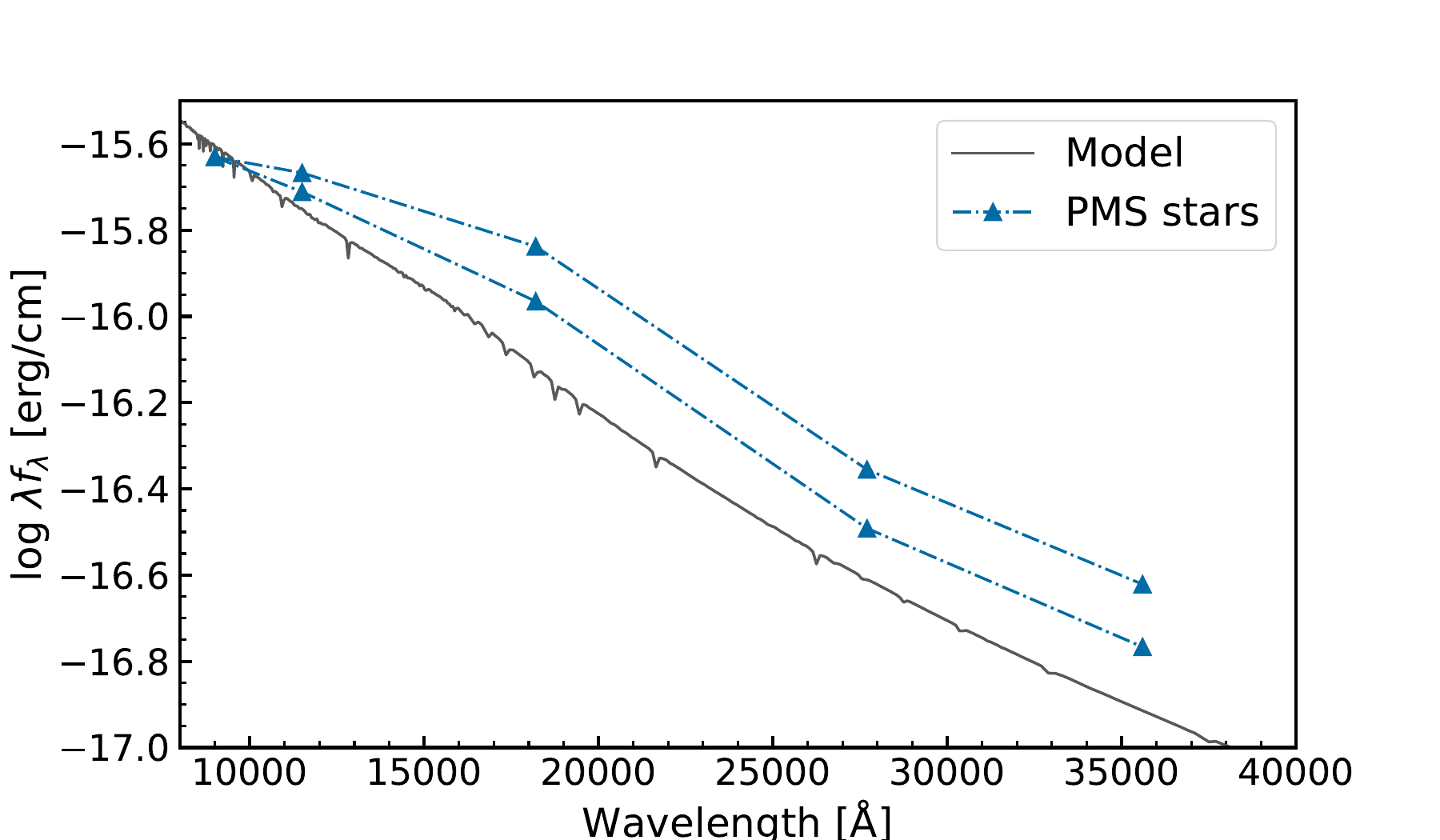}
     \caption{Model SED showing the logarithm of the flux density as a function of wavelength, zoomed in on the IR tail. Shown in blue are the two overlapping PMS stars and their flux density at different wavelengths. All curves are anchored to the model at 9000 $\mathrm{\AA}$}
         \label{fig:sed}
   \end{figure}

{Using all available JWST bands listed in Table \ref{tab:jwst}, we  {constructed} a spectral energy distribution (SED) for the two PMS stars in the overlapping region (Fig.\,\ref{fig:sed}). The SED traces the flux density as a function of wavelength and directly probes any near-IR excess above the stellar photosphere.} We  {converted} the magnitudes to flux density using
\begin{equation}
    f_{\nu} = 10^{-0.4(m-m_0)} \, \eta_{\rm MJSR} \, {BW}_{\rm eff} ,
\end{equation}
where $m$ is the measured magnitude, $m_0$ the zero-point (Vega) magnitude, $\eta_{\rm MJSR}$ the conversion factor (`\texttt{PHOTMJSR}') from counts per second to MJy\,sr$^{-1}$, and $BW_{\rm eff}$ the effective bandwidth of the respective filter. $\eta_{\rm MJSR}$ is taken from the header of the \texttt{.fits} files. We  {took} the literature values from the JWST documentation for the zero-point magnitudes \citep{zeropointjwst12023PASP..135d8001R, zeropointjwst22022AJ....163..267G} and effective bandwidths \citep{effBWjwst12008AJ....135.2245R, effBWjwst22005PASP..117..421T}. We note that the overlapping region falls entirely within the B1 detector on NIRCam; we thus adopted the values specifically for B1 (or B). They can be found in Table \ref{tab:jwst}.

\begin{table}[t]
\caption{NIRCam photometric band parameters. }
\centering
\begin{tabular}{llll}
\hline
\hline
{Band}                       & ~~~{$m_0$} & {$\eta_{\rm MJSR}$} & {${\rm BW_{\rm eff}}$} \\ \hline
{F090W} & 26.13         & 3.282             & 0.194            \\
{F115W} & 25.95         & 2.698             & 0.225            \\
{F182M} & 25.01         & 3.328             & 0.238            \\
{F187N} & 22.26         & 38.93             & 0.024            \\
{F277W} & 25.09         & 0.491             & 0.672            \\
{F356W} & 24.72         & 0.408             & 0.787            \\
{F405N} & 20.89         & 9.711             & 0.046            \\
{F430M} & 22.67         & 1.880             & 0.228           \\
\hline
\end{tabular}
\tablefoot{For each filter, the table provides the zero-point magnitude ($m_0$), the conversion factor from count s$^{-1}$ to MJy sr$^{-1}$ ($\eta_{\rm MJSR}$), and the effective bandwidth (BW$_{\rm eff}$).}
\label{tab:jwst}
\end{table}

Shown in Fig. \ref{fig:sed} are the SEDs for the two PMS stars in the overlapping region. These are compared to the model atmosphere of a MS 6500\,K star, again by \cite{bessel1998A&A...333..231B}. For the sake of comparison, the SEDs are normalised to this model continuum at 9000 $\mathrm{\AA}$, the lowest central wavelength in the JWST photometry. Compared to the model atmosphere for MS stars, the two PMS stars exhibit a moderate near-IR excess at all wavelengths. The observed excess is fully consistent with the one measured spectroscopically for PMS stars in the core of NGC\,346 by \cite{demarchietal2024}.  We note that the data point at $\sim 18700$\,\AA\ captures the flux density measured through the F187N narrow-band filter of NIRCam and includes not only the stellar continuum but also the Pa$\alpha$ emission line, which is the strongest H recombination line in this wavelength range and is connected with the accretion process. However, even if that data point is neglected, the SED of the stars is still consistent with the presence of a dusty disc.

{The detection of both H$\alpha$ excess (from our HST photometry) and near-IR excess (from JWST) for these two objects constitutes strong evidence for their PMS nature. Spectroscopic follow-up would allow a direct measurement of line strengths and profiles, to unambiguously confirm accretion activity and better constrain the disc structure.}

\section{Summary and conclusions}\label{sec:conclusion}

In this work, we studied the star formation in a field outside the cluster NGC 346 in the SMC {by analysing HST photometry in the $V$, $I$, and H$\alpha$ bands. A total of 12\,649 sources were selected as having reliable photometry with uncertainties $\sigma_{\rm H\alpha} \leq 0.3$ {mag} and $\sigma_{\rm VI} \leq 0.1$ {mag}}. We found a population of PMS stars and  {analysed} their distribution and physical properties{, studying the effect of young massive stars on the PMS population, as well as the proper motion of the stars in this region. Finally, we compared our results with JWST photometric observations of an overlapping region.} The main results of this  {work} can be summarised as follows:

\begin{enumerate}
\item 
 {The CMD shows} evidence of populations of roughly $10, 60, 400$ Myr, and 5 Gyr, which agrees with known star-forming episodes in the SMC.
\item 
 {We identified} 137 bona fide PMS stars with H$\alpha$ excess greater than 5$\sigma$ and $|W_{\rm eq}| \geq 20\,\mathrm{\AA}$  {in emission. The detected H$\alpha$ emission suggests the presence of an accreting PMS population in this low-density field, well outside the boundaries of the main NGC\,346 cluster.}
\item 
 {Most PMS stars are around 16\,Myr old, with $\langle M \rangle = 0.80 \pm 0.16$\,M$_\odot$.}
\item 
Dividing the PMS stars into two groups younger and older than 16 Myr, we find that the younger stars have a higher average mass, accretion luminosity, and mass accretion rate, as expected.
\item 
The PMS stars in the studied region have  {an overall} median mass accretion rate of $\dot{M}_{\rm acc} = (8 \pm 3) \times 10^{-9}$\,M$_\odot$\,yr$^{-1}$.  {This is comparable to other low-density star-forming regions in the SMC \citep{vlasblomdemarchi2023, tsiliaetal2023}, but lower than in the NGC\,346 cluster, where median values range from $(1.5 \pm 0.6) \times 10^{-8}$ to $(1.1 \pm 0.7) \times 10^{-7}$\,M$_\odot$\,yr$^{-1}$ depending on the age group \citep{PMSngc346guido2011ApJ...740...11D}.}
\item 
 {The PMS stars are distributed uniformly across the field without a significant spatial gradient; a local under-density is found in the southern portion of the field, coinciding with the concentration of massive stars.}
\item 
{The fraction} of PMS stars  {relative to} the total stellar population decreases closer to the barycentre of the young massive stars, and so does the mass accretion rate.  {This suggests an interaction between the PMS stars and the group of massive stars, which might accelerate disc dispersal.}
\item 
The massive stars, in the range $8-20$\,M$_\odot$, are four times more numerous than those in the ONC, and their number distribution agrees well with a standard Kroupa IMF.
\item 
 {The \textit{Gaia} DR3 proper motions of the massive stars are consistent with the bulk motion of the SMC, with no residual preferential direction, indicating that neither outflow nor rotation due to the nearby NGC\,346 cluster is detectable.}
\item 
 {Comparing HST to JWST photometry in the overlapping region, we find that two bona fide PMS stars show near-IR and Pa$\alpha$ excess, consistent with the presence of dusty circumstellar discs.}
\end{enumerate}

{With this work, we have} extended the knowledge of star formation in a seemingly inconspicuous region in the SMC, well outside the high-density active NGC\,346 star-forming region,  {where} metallicities are comparable to those prevailing in the early Universe. The physical properties of the population of PMS stars that we find appear to be affected by the ionising radiation of the coeval nearby massive stars.  {These results highlight the contribution of low-density field environments to the overall star formation budget, and the role that local massive stars play in shaping accretion processes in their vicinity.} Rather than being confined to dense clusters, massive stars in the field appear capable of shaping their surroundings,  {by} contributing meaningfully to galactic evolution and  {influencing local star formation}. As such, understanding the conditions that permit massive stars to form in isolation or sparse groups is essential for constructing more comprehensive models of stellar populations and determining their impact across different galactic contexts.

\section*{Data availability}

{Table\,\ref{tab:cds} is only available in electronic form at the CDS via anonymous ftp to cdsarc.u-strasbg.fr (130.79.128.5) or via http://cdsweb.u-strasbg.fr/cgi-bin/qcat?J/A+A/.}

\begin{acknowledgements}
We thank an anonymous referee for a careful reading of the manuscript and for constructive comments that helped us improve its clarity. We are grateful to Katja Fahrion for sharing with us her JWST/NIRCam photometry ahead of publication. This research is based on observations made with the NASA/ESA \textit{Hubble} Space Telescope obtained from the Space Telescope Science Institute, which is operated by the Association of Universities for Research in Astronomy, Inc., under NASA contract NAS 5–26555. These observations are associated with programmes 10248 and 13009.
This research has made use of Python, \url{https://www.python.org}, of NumPy \citep{harris2020array}, Astropy \citep{2013A&A...558A..33A}, and Matplotlib \citep{2007CSE.....9...90H}.
This work has made use of data from the European Space Agency (ESA) mission
{\it Gaia} (\url{https://www.cosmos.esa.int/gaia}), processed by the {\it Gaia} Data Processing and Analysis Consortium (DPAC, \url{https://www.cosmos.esa.int/web/gaia/dpac/consortium}). Funding for the DPAC has been provided by national institutions, in particular the institutions participating in the {\it Gaia} Multilateral Agreement.
This work is based in part on observations made with the NASA/ESA/CSA \textit{James Webb} Space Telescope and associated with programme 1227.
\end{acknowledgements}

\bibliographystyle{aa}
\bibliography{references}

\end{document}